\documentclass{PoS}

\usepackage{amssymb,amsfonts,amsmath,pifont,amscd}

\def\beq{\begin{equation}}
\def\eeq{\end{equation}}
\def\bea{\begin{align}}
\def\eea{\end{align}}
\def\Tr{{\rm Tr}}
\newcommand{\ud}{\mathrm{d}}

\renewcommand{\t}{\theta}

\newcommand{\om}{\omega}
\newcommand{\Om}{\Omega}
\renewcommand{\d}{\delta}
\newcommand{\D}{\Delta}

\newcommand{\s}{\sigma}
\newcommand{\vph}{\varphi}

\newcommand{\cd}{\mathcal{D}}
\newcommand{\bsp}{\boldsymbol{\varphi}}
\newcommand{\bspsi}{\boldsymbol{\psi}}
\newcommand{\prt}{\partial}
\newcommand{\bsD}{\boldsymbol{\Delta}}
\newcommand{\bsOm}{\boldsymbol{\Omega}}
\newcommand{\no}{\nonumber}
\newcommand{\non}{\nonumber \\}

\title{Lattice supersymmetry in 1D with two supercharges}

\ShortTitle{Lattice supersymmetry in 1D with two supercharges}

\author{\speaker{Sergio Arianos}, \speaker{Alessandro D'Adda} \\
        INFN, Torino\\
        E-mail: \email{arianos@to.infn.it}, \email{dadda@to.infn.it}}

\author{Noboru Kawamoto, Jun Saito\\
        Department of Physics, Hokkaido University\\
        E-mail: \email{kawamoto@particle.sci.hokudai.ac.jp}, 
\email{saito@particle.sci.hokudai.ac.jp}}

\abstract{A consistent formulation of a fully supersymmetric
theory on the lattice has been a long standing challenge. In
recent years there has been a renewed interest on this problem
with different approaches. At the basis of the formulation we
present in the following there is the Dirac-K\"ahler twisting
procedure, which was proposed in the continuum for a number of
theories, including $N=4$ SUSY in four dimensions. Following the
formalism developed in recent papers, an exact supersymmetric
theory with two supercharges on a one dimensional lattice is
realized using a matrix-based model. The matrix structure is
obtained from the shift and clock matrices used in two dimensional
 non-commutative field theories. The matrix structure reproduces on a one
 dimensional lattice the expected modified Leibniz rule. Recent claims of
 inconsistency of the formalism are discussed and shown not to be relevant.
}

\FullConference{The XXV International Symposium on Lattice Field Theory\\
                 July 30 - August 4 2007\\
                 Regensburg, Germany}

\begin{document}

\section{Introduction}

A consistent formulation of supersymmetry on the lattice is a long
standing problem. A number of approaches that allow to preserve
exactly one supersymmetry on the lattice in theories with an
extended supersymmetry have been proposed in recent years
\cite{deconstruction,twist-lattsusy,Sugino}. A more ambitious approach,
aiming to preserve exactly all supersymmetries in some extended
supersymmetric model, was also proposed
\cite{D'Adda:2004jb,D'Adda:2005zk,DKKN3}.

Like for many of the previous formulations one of the key
ingredients of this approach is the use of Dirac-K\"ahler fermions
on the lattice to overcome the doubling problem. Its main feature
however is the use of an extended lattice where the standard
links, corresponding to the discrete elementary translations on
the lattice, are implemented by "fermionic" links that correspond
to the action of supersymmetry charges. The need for these extra
links was the result of a careful analysis of the modified
"shifted" Leibniz rules, that both translations and supersymmetry
transformation have for consistency to satisfy on a lattice when
acting on a product of (super)fields. The structure of the
extended lattice thus reflects the structure of the supersymmetry
algebra, and a consistent solution is found only for some specific
extended superalgebra like the $N=2$ superalgebra for $D=2$
\cite{D'Adda:2004jb,D'Adda:2005zk}, the $N=4$ 
superalgebra for $D=3$\cite{DKKN3}, and
the $N=4$ superalgebra for $D=4$ \cite{D'Adda:2005zk}. Consistently
the supersymmetry charges are in this approach associated to
links, rather than to sites. In connection with this point a
number of criticisms were put forward
\cite{Bruckmann:2006ub,Bruckmann:2006kb}, with the claim that the
link nature of supercharges leads to inconsistencies and
ambiguities in the definition of the supersymmetry
transformations. Very recently \cite{Damgaard} it was shown that
this approach fits within the scheme of Kaplan's orbifold
formulation \cite{deconstruction} and the actual invariance of the action
proposed in \cite{D'Adda:2005zk} under all Susy charges has been
questioned.

In order to investigate the above issues we considered a simple
one dimensional supersymmetric model with two supercharges, the
same model already considered by Bruckmann and de Kok in
\cite{Bruckmann:2006ub}. A detailed analysis of this model on the
lattice will be the subject of a future publication \cite{ADKS}
and some of the main results are anticipated in the second part of
the present report. In particular it is shown that supersymmetry
transformations are consistently and unambiguously determined on
the lattice by using the superfield formalism. In the first part
of the report we discuss some general features of (one
dimensional) lattice theories (not necessarily supersymmetric). In
particular we shall study the connection between the modified
Leibniz rule mentioned above and the translational invariance of
the action on the lattice and show how they follow from
dimensional reduction of a two dimensional theory  formulated on a
non-commutative lattice.

\section{One dimensional lattice models }

\subsection{Matrix representation and modified Leibniz rules}

Consider a one dimensional lattice with $N$ sites and periodic
boundary conditions and a scalar field $\varphi$, defined on the sites of the
lattice. Let $\varphi_r$ ($r=1,2, \cdots , N$) be the value of the field $\vph$
on the $r$th site of the lattice.
The $N$ numbers $\vph_r$ can be regarded as the eigenvalues of an $N \times N$
diagonal matrix $\boldsymbol{\varphi}$:
\beq
\bsp = \begin{pmatrix}
\vph_1 & 0 & 0 & 0 & \cdots & 0 \\
0 & \vph_2 & 0 & 0 & \cdots & 0 \\
0 & 0 & \vph_3 & 0 & \cdots & 0 \\
\vdots & \vdots & \vdots & \vdots & \ddots & \vdots \\
0 & 0 & 0 & 0 & \cdots & \vph_N
\end{pmatrix} \label{matrixfield}
\eeq
whose rows and columns  are in one to one correspondence with
the sites of the lattice. Notice that the ordering of the rows and
columns is the same as the one of the lattice sites, so that
neighboring eigenvalues correspond to the values of the field in
neighboring sites. Derivatives are replaced on the lattice by
finite differences:
\beq (\Delta_{+}\varphi)_r = \varphi_{r+1} -
\varphi_r \label{finitediff}. \eeq
 The breaking of the
translational invariance due to the discrete nature of the lattice
results into a violation of the Leibniz rule when the finite
difference of a product of two functions is considered. As
discussed in detail in ref.\ \cite{D'Adda:2004jb}, to which we
refer for a more exhaustive treatment, a modified Leibniz rule
holds in place of the usual one:
 \beq
 (\Delta_{+}\varphi \psi)_r =
(\Delta_{+}\varphi)_r \psi_r + \varphi_{r+1} (\Delta_{+}\psi)_r.
\label{modleib1}
 \eeq
In matrix notation finite differences may be represented using the
shift matrices $\bsD_+$ and $\bsD_-$:
\beq \bsD_+= \begin{pmatrix}
0 & 1 & 0 & 0 & \cdots & 0 \\
0 & 0 & 1 & 0 & \cdots & 0 \\
0 & 0 & 0 & 1 & \cdots & 0 \\
\vdots & \vdots & \vdots & \vdots & \ddots & \vdots \\
0 & 0 & 0 & 0 & \cdots & 1 \\
1 & 0 & 0 & 0 & \cdots & 0
\end{pmatrix}\;,
\qquad \bsD_-=\bsD_+^{-1}= \begin{pmatrix}
0 & 0 & 0 & \cdots & 0 & 1 \\
1 & 0 & 0 & \cdots & 0 & 0 \\
0 & 1 & 0 & \cdots & 0 & 0 \\
0 & 0 & 1 & \cdots & 0 & 0 \\
\vdots & \vdots & \vdots & \ddots & \vdots & \vdots \\
0 & 0 & 0 & \cdots & 1 & 0
\end{pmatrix}
\label{Delta}
\eeq
namely, in components
\beq
(\bsD_+)_{rs}=\d_{r,s-1}\;, \qquad (\bsD_-)_{rs}=\d_{r,s+1}\;.
\label{Deltacomp}
\eeq

In the continuum the derivative $\partial \varphi$ is just the
commutator $[\partial,\varphi]$; on the lattice however the
commutator $[ \bsD_+ , \bsp ]$ is not diagonal, its non vanishing
matrix elements being on a shifted diagonal as in $\bsD_+$. In
order to write the finite difference (\ref{finitediff}) as a
function defined on the lattice sites, namely a diagonal matrix,
we have to define it as: \beq (\bsD_+ \bsp) = -\bsD_+ [ \bsD_- ,
\bsp ]= \begin{pmatrix}
\vph_2-\vph_1 & 0 & 0 & 0 & \cdots & 0 \\
0 & \vph_3-\vph_2 & 0 & 0 & \cdots & 0 \\
0 & 0 & \vph_4-\vph_3 & 0 & \cdots & 0 \\
\vdots & \vdots & \vdots & \vdots & \ddots & \vdots \\
0 & 0 & 0 & 0 & \cdots & \vph_1-\vph_{N}
\end{pmatrix}\label{findiff} \eeq
The factor $\bsD_+$ in front of the commutator is responsible for
the violation of the Leibniz rule, in fact we have:
 \beq (\bsD_+
\bsp \bspsi) = (\bsD_+ \bsp ) \bspsi + \bsD_+ \bsp \bsD_- (\bsD_+
\bspsi ) \label{modleib2} \eeq which is completely equivalent to
(\ref{modleib1}). Notice that $\bsD_+ \bsp \bsD_-$ is a "shifted"
field, where the eigenvalue $\varphi_r$ has been replaced by
$\varphi_{r+1}$: $(\bsD_+ \bsp \bsD_-)_r = \bsp_{r+1}$. The
modified Leibniz rule (\ref{modleib2}) reflects the fact that
translational symmetry on the lattice is a discrete, and not a
continuous symmetry. To make this point clear consider an action
given as the trace of a product of fields $\bsp_i$: \beq S = \Tr~~
\bsp_1 \bsp_2 \cdots \bsp_r \label{traction} \eeq The trace
corresponds to the sum over all lattice sites, and translational
invariance can be simply expressed as the invariance of
(\ref{traction}) under \beq \bsp_i \rightarrow \bsD_+ \bsp_i
\bsD_- = \bsp_i + \delta \bsp_i \label{transl} \eeq where $\delta
\bsp_i  = (\bsD_+ \bsp)$ as defined in (\ref{findiff}). When the
r.h.s.\ of (\ref{transl}) is inserted into (\ref{traction}) all
orders of $\delta \bsp_i$ must be kept in order to preserve the
exact symmetry and the variation of the Lagrangian can be cast in
the form:
\begin{align}
\delta & \left( \bsp_1 \bsp_2 \cdots \bsp_r \right) = (\delta \bsp_1) \bsp_2
\cdots \bsp_{r-1} \bsp_r + (\bsp_1 + \delta \bsp_1) 
(\delta \bsp_2) \bsp_3 \cdots \bsp_r +
\nonumber \cdots \\ + &  (\bsp_1 + \delta \bsp_1)(\bsp_2 + \delta \bsp_2) \cdots
(\bsp_{r-1} + \delta \bsp_{r-1})(\delta \bsp_r) \label{modleib3}
\end{align}
which is again the modified Leibniz rule. It is clear that
eq.~(\ref{modleib3}) follows directly from the transformation
(\ref{transl}) by keeping all orders in $ \delta \bsp_i$ while
linear terms in $\delta \bsp_i$ give the ordinary Leibniz rule
typical of the continuum limit. We stressed this point because the
situation is different in the supersymmetric theory discussed in the following
section: supersymmetry charges are non diagonal and hence supersymmetry
transformations of a product of superfields obey a modified
Leibniz rule on the lattice, as discussed in section 4 as
well as in previous papers~\cite{D'Adda:2004jb}, however these modified
Leibniz rules cannot be derived, at least in the present formulation, from a
field transformation as in (\ref{transl}).

\subsection{Dimensional reduction and non-commutative lattice}

In the matrix representation of fields (\ref{matrixfield}) there
is a one-to-one correspondence between rows (or columns) and
lattice points. If the lattice contains $N$ points, a generic
matrix will have $N^2$ matrix elements and each matrix element
$\varphi_{ij}$ is associated to an ordered link joining two
arbitrary points $i$ and $j$ of the lattice. A generic matrix then
describes a completely non-local object on the lattice. In order
to recover the lattice structure restriction must be imposed on
the matrix $\bsp$ for it to describe a local or almost local
field.

Such restrictions are just the analogue, in this simple one
dimensional model, of the orbifold conditions used by Kaplan and
collaborators in their approach to lattice supersymmetry
~\cite{deconstruction}. In fact they can be expressed, as in
~\cite{deconstruction}, in terms of the "clock" matrix $\Om$ defined by~\footnote{In the definition of $\om$ we adopt the same notation used in \cite{Bars:1999av}.}:
\beq \bsOm =
\begin{pmatrix}
1 & 0 & 0 & \cdots & 0 \\
0 & \om & 0 & \cdots & 0 \\
0 & 0 & \om^2 & \cdots & 0 \\
\vdots & \vdots & \vdots & \ddots & \vdots \\
0 & 0 & 0 & \cdots & \om^{N-1}
\end{pmatrix},\qquad
\begin{aligned}
\om& =e^{\frac{-i2\pi}{N}}, \\
\om^N& =1.
\end{aligned}
\label{clock} \eeq

A field defined on the lattice sites and described by a diagonal
matrix $\bsp$ simply commutes with $\Om$, but in general we shall
be interested in fields defined on the lattice links (like gauge
fields) whose matrix representation is of the form \beq
\bsp^{(a)}_{ij}= \delta_{i+a,j}\varphi^{(a)}_i \label{shiftdiag}
\eeq where $a=\pm 1$ for the link variables while higher values of
$a$ denote more non-local fields, like the ones involving higher
derivatives. It is easy to check that requiring that the field $
\bsp^{(a)}$ is of the form (\ref{shiftdiag}) is the same as
imposing the following orbifold condition:
 \beq \bsp^{(a)} \Om
= \om^a \Om \bsp^{(a)}. \label{orbifold} \eeq In particular, as
the finite difference operator $\bsD_+$ and its conjugate
$\bsD_-$ are such matrices with $a=1$ and $a=-1$ respectively, we
also have: \beq \bsD_{\pm}\Om = \om^{\pm 1} \Om \bsD_{\pm}.
\label{noncomm} \eeq

The orbifold conditions (\ref{orbifold}) have an interesting
interpretation in terms of non-commutative geometry.
Non-commutative geometry on a discrete periodic lattice has been
studied by several authors ~\cite{nbi-noncom,Bars:1999av}.
We adopt the approach of Bars and Minic which is the most
convenient for our purpose. The first thing to notice is that
before the orbifold conditions are applied a field $\bsp$ is
represented by an $N \times N$ matrix, namely it contains $N^2$
degrees of freedom, that is exactly the number of degrees of
freedom of a scalar field on a two dimensional lattice with $N$
links in each direction. So the reduction of the degrees of
freedom from $N^2$ to $N$ amounts effectively to some kind of
dimensional reduction. In order to understand what type of
dimensional reduction that is, let us consider a non-commutative
two dimensional lattice as defined in ref.~\cite{Bars:1999av}.
The lattice structure is imposed by requiring that the two
coordinates $X_1$ and $X_2$ are operators with $N$ discrete
eigenvalues $(x_i)_n= n a$ where $a$ is the lattice spacing and
$n$ an integer modulo $N$. The non-commutativity of the
coordinates $X_1$ and $X_2$ is given by: \beq \left[
\frac{X_1}{L}, \frac{X_2}{L} \right] = \frac{i}{2 \pi N b},
\label{nc} \eeq where $b$ is an arbitrary integer and $L = N a$ is
in each direction the size of the "box" with periodic boundary
conditions. The quantisation of $b$, as explained in
~\cite{Bars:1999av}, is a consequence of the periodic boundary
conditions and of the discreteness of the spectrum of eigenvalues.
The normalization of the coordinates in (\ref{nc}) has been chosen
 to show that if the continuum limit is done keeping the size $L$
of the box, namely the infrared cutoff, fixed the r.h.s.\ vanishes
in that limit, as continuum limit and large $N$ limit coincide in
this case. We shall choose here for convenience $b=1$. It is shown
in ~\cite{Bars:1999av} that the shift operator $\bsD_+$ and the
clock operator $\Om$ are the translation operators of one lattice
unit in the two lattice directions.  In the base where $X_1$ is
diagonal the shift operator $\bsD_+$ is
 translation operator of one lattice unit along the positive $X_1$
direction ( and its conjugate $\bsD_-$ in the negative direction);
 $\Om$ on the other hand is the translation operator of one lattice unit
 along $X_2$.
For $b=1$ the translation operators can be written in terms of the
coordinate operators as: \beq \bsD_+ = \exp{i\frac{2 \pi
}{N} \frac{X_2}{a}}= \om^{-\hat{X}_2}, \qquad\Om =
\exp{-i\frac{2 \pi }{N} \frac{X_1}{a}}=\om^{\hat{X}_1}
\label{transop} \eeq where $\hat{X}_i = \frac{X_i}{a}$ are the
coordinates normalized to the lattice spacing (the eigenvalues of
$\hat{X}_i$ are integers) and $\om$ the $N$th root of the
identity, as defined above. The orbifold conditions
(\ref{orbifold}) have now a clear interpretation as they represent
different types of compactification along the $X_2$ direction. In
the simplest case of a scalar field the condition $[\bsp , \Om]=0$
simply states that there is no dependence on the $X_2$ direction.
However the non-commutativity (\ref{noncomm}) of $\Om$ and
$\bsD_+$ implies that $\bsp$ and $[\bsD_+, \bsp]$ do not obey the
same orbifold/dimensional-reduction conditions, hence the
impossibility of defining the finite difference operation as a
commutator and the need for a modified Leibniz rule. A more
precise understanding of this is obtained, always following
ref.~\cite{Bars:1999av}, by introducing the analogue on a two
dimensional non-commutative lattice of the Moyal product. Let us
introduce with~\cite{Bars:1999av} the most general translation
operator: \beq \hat{v}_p = \exp{i p_{\mu} X_{\mu}}
\label{gtras} \eeq with $p_{\mu}$ the discretized momenta on the
lattice: \beq p_1 = -\frac{2 \pi k_2}{a N}, \qquad p_2 =
\frac{2 \pi k_1}{a N},\qquad k_1,k_2 \ \rm{integers}.
\label{momenta} \eeq By using the relations (\ref{transop}) and
the Baker-Hausdorff formula  $\hat{v}_p$ can be written as: \beq
\hat{v}_p\equiv \hat{v}_{k_1,k_2} = \om^{\frac{k_1 k_2}{2}}
\Om^{k_2} {\bsD_+}^{k_1}. \label{gtras2} \eeq Notice that because
of the Baker-Hausdorff term $\om^{\frac{k_1 k_2}{2}}$
$\hat{v}_{k_1,k_2}$ is  periodic with period $ 2 N $ (and not $N$)
in $k_1$ and $k_2$,  and hence we shall take $k_1$ and $k_2$ to be
integers modulo $2 N$. Given a scalar field $\varphi$ represented
by an $N \times N$ matrix $\bsp$ we can define its representation
in the space of the discrete momenta $k_1$ and $k_2$ as: \beq
\hat{\bsp}_{k_1,k_2} = \frac{1}{N} \Tr\,\hat{v}_{k_1,k_2} \bsp
\label{momrep} \eeq where, due to the $\om^{\frac{k_1 k_2}{2}}$
factor in (\ref{gtras2}), $k_1$ and $k_2$ are defined modulo $2 N$
with the symmetry: \beq \hat{\bsp}_{k_1+N,k_2}= (-1)^{k_2}
\hat{\bsp}_{k_1,k_2}\qquad {\rm and}\qquad \hat{\bsp}_{k_1,k_2+N}=
(-1)^{k_1} \hat{\bsp}_{k_1,k_2}. \label{symimp} \eeq The coordinate
representation of $\bsp$ can be obtained by doing a discrete
Fourier transform of $\hat{\bsp}_{k_1,k_2}$: \beq \bsp(\xi)=
\frac{1}{N} \sum_{k_1,k_2}\om^{k_1 \hat{\xi}_2 - k_2 \hat{\xi}_1}
\hat{\bsp}_{k_1,k_2}= \frac{1}{N}\Tr\,\hat{\Delta}(\xi)\bsp
\label{coordrep} \eeq where we have introduced the matrix
$\hat{\Delta}(\xi)$ defined as: \beq \hat{\Delta}(\xi) =
\frac{1}{N} \sum_p \exp i p_{\mu} \left( X_{\mu} - \xi_{\mu}
\right) = \frac{1}{N} \sum_{k_1,k_2} \om^{\frac{k_1 k_2}{2}}
\Om^{k_2}{\bsD_+}^{k_1}\om^{k_1 \hat{\xi}_2 - k_2 \hat{\xi}_1}
\label{deltahat}. \eeq As before $\hat{\xi}_i = \frac{\xi_i}{a}$
are normalized to the lattice spacing and the sum over $k_i$ goes
from $1$ to $2 N$. With $k_1$ and $k_2$ defined modulo $2 N$ the
lattice positions $\hat{\xi}_i$ can take both integer and
half-integer values (modulo $N$), with the integer values coming
(thanks to the symmetry (\ref{symimp})) from the even values of
$k_i$ and the half-integer values from the odd values of $k_i$.
The matrix $\hat{\Delta}(\xi)$ provides through eq.~(\ref{coordrep})
the map between the matrix representation of the
field $\bsp$ and its lattice coordinate representation.
Eq.~(\ref{coordrep}) can be inverted: \beq \bsp = \sum_{\xi_{\mu}}
\bsp(\xi) \hat{\Delta}(\xi). \label{invert} \eeq The
non-commutativity of the coordinates has as a consequence that the
product of fields is not local in the coordinate representation:
it is the lattice analogue of the Moyal product. It is defined, as
in ~\cite{Bars:1999av}, by the relation: \beq \bsp_1 \diamond
\bsp_2 (\xi) = N^{-1} \Tr\,\bsp_1 \bsp_2 \hat{\Delta}(\xi) = N^{-1}
\sum_{\hat{x},\hat{y}}
 \om^{-2\epsilon^{\mu\nu}(\hat{x}_\mu - \hat{\xi}_\mu)(\hat{y}_\nu
- \hat{\xi}_\nu)} \bsp_1(\hat{x})\bsp_2(\hat{y})
\label{diaproduct} \eeq and denoted as "diamond" product. Further
properties of the diamond product can be found in
~\cite{Bars:1999av}, our aim here is to study what it becomes when
the orbifold (dimensional reduction) conditions (\ref{orbifold})
are imposed. Let us then consider a field $\bsp^{(a)}$ of the
shifted diagonal form (\ref{shiftdiag}), namely satisfying the
orbifold condition (\ref{orbifold}). By using the explicit
expression of $\hat{\Delta}(\xi)$ given in (\ref{deltahat}) one
can easily find the coordinate representation of the field
$\bsp^{(a)}$: \beq 
\varphi^{(a)}(\xi) 
=\frac{1}{N}\omega^{-a\hat{\xi}_2}\varphi_{\hat{\xi}_1+\frac{a}{2}-1}^{(a)}.
\label{coorph} \eeq As
a result of the dimensional reduction the dependence on
$\hat{\xi}_2$ in (\ref{coorph}) is trivial, with the shift $a$
interpreted as a constant momentum in the compactified direction
$\xi_2$. The index $\hat{\xi}_1 - \frac{a}{2}$ in (\ref{coorph})
should be an integer, so fields with even shifts $a$ are defined
on integers values of $\hat{\xi}_1$ while fields with odd shift
$a$ are defined on half-integer values of $\hat{\xi}_1$. Consider
now two shifted diagonal fields $\bsp_1^{(a_1)}$ and $\bsp_2^{(a_2)}$.
In the matrix representation the product of the two fields is a
shifted diagonal matrix with shift equal to $a_1+a_2$. In the
coordinate representation the diamond product of the two fields
can be easily calculated and is given by: \beq
\varphi_1^{(a_1)}\diamond \varphi_2^{(a_2)}(\xi) =N
\varphi_1^{(a_1)}(\hat{\xi}_1-\frac{a_2}{2},\hat{\xi}_2)
\varphi_2^{(a_2)}(\hat{\xi}_2+\frac{a_1}{2},\hat{\xi}_2).
\label{dprod} \eeq Given the trivial dependence on $\hat{\xi}_2$
of both terms, the  $\hat{\xi}_2$ dependence of the r.h.s.\ is just
$\om^{- (a_1+a_2) \hat{\xi}_2}$, namely the product field has a
shift (i.e. momentum in $\xi_2$ direction) equal to $a_1+a_2$.
Eq.~(\ref{dprod})  shows that the diamond product is non-commutative
also the dimensional reduction conditions have been imposed. This
is due to the shifts in the $\hat{\xi}$ dependence at the r.h.s.\ 
of (\ref{dprod}). This type of "mild" non-commutativity is just
the one introduced in ~\cite{D'Adda:2004jb} in order to have all
supersymmetries exactly preserved on the lattice. It can also be
interpreted in terms of link variables according to the scheme
developed in ~\cite{D'Adda:2005zk} for supersymmetric lattice
gauge theories. In fact if we consider
$\varphi^{(a)}(\hat{\xi}_1,\hat{\xi}_2)$ as a degree of freedom
associated in the $\xi_1$ space to the link of length $a$
$(\hat{\xi}_1-\frac{a}{2},\hat{\xi}_1+\frac{a}{2})$, the diamond
product (\ref{dprod}) can be interpreted as the product of two
successive link variable of length $a_1$ and $a_2$ starting in
$\hat{\xi}_1-\frac{a_1+a_2}{2}$ and ending in
$\hat{\xi}_1+\frac{a_1+a_2}{2}$. \newpage

\section{The $N=2$ supersymmetric model in one dimension}
\subsection{Matrix representation of a Grassmann algebra}

In order to define the $N=2$ supersymmetric quantum mechanics on a one
dimensional lattice it is convenient to introduce a matrix representation
for the two Grassmann variables $\t_1$ and $\t_2$

\begin{align}
\t_1&\equiv\s_+ \otimes \mathbf{1} \otimes \D_+\;,  & \t_2&\equiv\s_3 
\otimes \s_+ \otimes \D_-\;, \\
\frac{\prt}{\prt\t_1}&\equiv\s_- \otimes \mathbf{1} \otimes \D_-\;, &
\frac{\prt}{\prt\t_2}&\equiv\s_3 \otimes \s_- \otimes \D_+\;;
\end{align}
or explicitly
\begin{align}
\t_1&\equiv \begin{pmatrix}
0 & 0 & \D_+ & 0 \\
0 & 0 & 0 & \D_+ \\
0 & 0 & 0 & 0 \\
0 & 0 & 0 & 0
\end{pmatrix}
& \t_2&\equiv \begin{pmatrix}
0 & \D_- & 0 & 0 \\
0 & 0 & 0 & 0 \\
0 & 0 & 0 & -\D_- \\
0 & 0 & 0 & 0
\end{pmatrix} \label{theta}\\
& & & \non \frac{\prt}{\prt\t_1}&\equiv \begin{pmatrix}
0 & 0 & 0 & 0 \\
0 & 0 & 0 & 0 \\
\D_- & 0 & 0 & 0 \\
0 & \D_- & 0 & 0
\end{pmatrix}
& \frac{\prt}{\prt\t_2}&\equiv \begin{pmatrix}
0 & 0 & 0 & 0 \\
\D_+ & 0 & 0 & 0 \\
0 & 0 & 0 & 0 \\
0 & 0 & -\D_+ & 0
\end{pmatrix} \label{dtheta}
\end{align}
where the entries of the above matrices are $N\times N$ matrices
and $\D_+$ and $\D_-$ are the shift matrices defined in (\ref{Delta}).

It is straightforward to check that the matrices (\ref{theta}) and
(\ref{dtheta}) satisfy the standard Grassmann algebra of the
$\theta$ variables. This matrix representation is quite general
and can be easily extended to an arbitrary number $n$ of variables
by using direct products of $n$ Pauli matrices, namely $2^n \times
2^n$ matrices. Notice also that according to the approach of ref
~\cite{D'Adda:2004jb} both $\t_i$ and $\prt\t_i$ contain a shift operator
$\D_+$ (resp.\ $\D_-$) of one lattice unit, implying that ordinary
derivative will correspond to a shift of two lattice spacings.

\subsection{Fields and superfields}
The next ingredient we need in order to construct a supersymmetric 
lattice theory is a matrix representation of the fields. As usual 
we deal with bosonic fields, fermionic fields and superfields,
defined as follows.
\begin{itemize}
\item Bosonic field: a field which commutes with all $\t$'s and 
$\frac{\prt}{\prt\t}$'s. A straightforward calculation gives
\begin{align} \label{phi}
\hat{\vph}\equiv \begin{pmatrix}
\vph & 0 & 0 & 0 \\
0 & \D_+\vph\D_- & 0 & 0 \\
0 & 0 & \D_-\vph\D_+ & 0 \\
0 & 0 & 0 & \vph
\end{pmatrix}
&&
\begin{aligned}
\vph& \equiv N\times N ~~ \text{matrix} \\
\t_i\hat{\vph}& =\hat{\vph}\t_i \\
\frac{\prt}{\prt\t_i}\hat{\vph}& =\hat{\vph}\frac{\prt}{\prt\t_i}
\end{aligned}
\end{align}

\item Fermionic field: a field which anticommutes with all $\t$'s 
and $\frac{\prt}{\prt\t}$'s. A straightforward calculation gives
\begin{align} \label{psi}
\hat{\psi}\equiv \begin{pmatrix}
\psi & 0 & 0 & 0 \\
0 & -\D_+\psi\D_- & 0 & 0 \\
0 & 0 & -\D_-\psi\D_+ & 0 \\
0 & 0 & 0 & \psi
\end{pmatrix}
&&
\begin{aligned}
\psi& \equiv N\times N ~~ \text{fermionic matrix} \\
\t_i\hat{\psi}& =-\hat{\psi}\t_i \\
\frac{\prt}{\prt\t_i}\hat{\psi}& =-\hat{\psi}\frac{\prt}{\prt\t_i}
\end{aligned}
\end{align}
\item Superfield: a field which commutes with all $\t$'s but not 
with $\frac{\prt}{\prt\t}$'s. It has a standard expansion in powers of $\t$'s:
\beq
\Phi=\hat{\vph}+\t_1\hat{\psi}_1+\t_2\hat{\psi}_2+\t_1\t_2\hat{D}\;.
\eeq
In our matrix representation it can be written as
\beq
\Phi=\begin{pmatrix} \label{PHI}
\vph & -\psi_2\D_- & -\psi_1\D_+ & -D \\
0 & \D_+\vph\D_- & 0 & \D_+\psi_1 \\
0 & 0 & \D_-\vph\D_+ & -\D_-\psi_2 \\
0 & 0 & 0 & \vph
\end{pmatrix}.
\eeq
\end{itemize}
From equations (\ref{phi}, \ref{psi}, \ref{PHI}) it is apparent that the actual
building blocks of the model are the matrices ``without hat''
$\vph,\;\psi_1,\;\psi_2$ and $D$. Indeed these are the matrices which will be
identified with the usual fields in the continuum limit.

So far $\vph,\;\psi_1,\;\psi_2$ and $D$ are arbitrary $N\times N$ matrices with
$N^2$ degrees of freedom. In order to describe a one dimensional lattice of
size $N$ we need to apply to $\Phi$ an orbifold condition as in
(\ref{orbifold}), namely:

\beq \label{sorbifold}
[\hat{\Om},~ \Phi]=0\;,
\eeq
where $\hat{\Om}$ is the block diagonal matrix 
$\hat{\Om} = \mathbf{1} \otimes \mathbf{1}
\otimes \Om$. This orbifold condition may also be interpreted as a dimensional
reduction from a two-dimensional non-commutative lattice, but we will not
discuss this possibility here.

In terms of the component fields eq.(\ref{orbifold}) requires:
\begin{align}
[\Om,~ \vph]=[\Om,~ D]& =0 \non
\Om\psi_1-\om\psi_1\Om& =0 \label{orbcond} \\
\Om\psi_2-\om^{-1}\psi_2\Om& =0 \no
\end{align}
This means that $\vph$ and $D$ are diagonal matrices; $\psi_1$
(like $\D_-$) has non vanishing elements only on the one-down
diagonal; $\psi_2$ (like $\D_+$) has non vanishing elements only
on the one-up diagonal. Notice however that in the block matrix representation
(\ref{PHI}) all entries are diagonal $N \times N$ matrices.

\subsection{Supercharges and susy transformations}

The two supercharges of the $N=2$ supersymmetric quantum mechanics are given in
the continuum theory by
\beq
 Q_i=\frac{\prt}{\prt\t_i}+\t_i\frac{\prt}{\prt t}\;.
 \label{contcharges}
 \eeq
As we assigned to $\t_i$ a shift operator corresponding to one lattice unit, the
time derivative $\prt t$ will be associated on the lattice to the two units
shift operator $ \D_{\pm}^2$.
The correspondence between continuum and lattice operators will then be
\beq
\prt t \rightarrow  N \D_{\pm}^2
\label{drv}
\eeq
where the factor $N$ is needed to recover the continuum limit. In fact
$[ \D_{\pm}^2, \vph ]$ is of order of the lattice spacing
$ a=\frac{L}{N}$, namely, if the size $L$ of the lattice is kept fixed, of order
$1/N$.
In our lattice formulation, both the $\t$'s and
$\frac{\prt}{\prt\t}$ carry a shift. On the other hand it is necessary for
consistency that the two terms in $Q_i$ carry the same shift, and this
determines the supercharges on the lattice without ambiguity:

\begin{align} \label{superc}
&& Q_1=\frac{\prt}{\prt\t_1}+N \t_1\hat{\D}_-^2= \begin{pmatrix}
0 & 0 & N \D_- & 0 \\
0 & 0 & 0 &N \D_- \\
\D_- & 0 & 0 & 0 \\
0 & \D_- & 0 & 0
\end{pmatrix},
&&  Q_2=\frac{\prt}{\prt\t_2}+N \t_2\hat{\D}_+^2= \begin{pmatrix}
0 & N \D_+ & 0 & 0 \\
\D_+ & 0& 0 & 0 \\
0 & 0 & 0 & -N \D_+\\
0 & 0 & -\D_+ & 0
\end{pmatrix}.
\end{align}
$Q_1$ and $Q_2$ defined in (\ref{superc}) satisfy  the algebra of
supersymmetric quantum mechanics written in Majorana representation (see
\cite{Bruckmann:2006ub} and references therein) namely:
\begin{align}
 Q_1^2& = N \hat{\D}_-^2\;, & Q_2^2& = N \hat{\D}_+^2\;, \label{2lat} \\
\{Q_1,~Q_2\}& =0\;, & [Q_{1,\,2},~\D_{\pm}]& =0\;.
\end{align}

Supersymmetry transformations are naively obtained by taking the commutator 
of $Q_1$ and $Q_2$ with $\Phi$. However, for consistency we want the 
supersymmetry variations of $\Phi$ to commute with $\Om$, just like $\Phi$, 
and also to commute with all the $\t$'s. This is
obtained by defining
\begin{align} \label{d1d2}
\d_1\Phi=\hat{\eta}_1\hat{\D}_+[Q_1,~\Phi]\;, &&
\d_2\Phi=\hat{\eta}_2\hat{\D}_-[Q_2,~\Phi]\;,
\end{align}
where
\begin{align} \label{eta}
\hat{\eta}_1= \eta_1\begin{pmatrix}
\mathbf{1} & 0 & 0 & 0 \\
0 & -\mathbf{1} & 0 & 0 \\
0 & 0 & -\mathbf{1} & 0 \\
0 & 0 & 0 & \mathbf{1}
\end{pmatrix},
&& \hat{\eta}_2= \eta_2\begin{pmatrix}
\mathbf{1} & 0 & 0 & 0 \\
0 & -\mathbf{1} & 0 & 0 \\
0 & 0 & -\mathbf{1} & 0 \\
0 & 0 & 0 & \mathbf{1}
\end{pmatrix}
\end{align}
and $\eta_1$, $\eta_2$ are odd Grassmann parameters:
\begin{align}
\eta_i\psi_j=-\psi_j\eta_i && \forall~i,\,j=1,\,2\;.
\end{align}
The matrix in (\ref{eta}) anticommutes with all $\t$'s, so that $\hat{\eta}_i$ 
anticommutes with both $\t$'s and the fermionic fields.

By doing explicit matrix computations we obtain, in terms of component fields
\begin{align}
\d_1\vph& =\eta_1\D_+\psi_1\;, &  \d_1D& =\eta_1 N\D_+[\D_-^2,~\psi_2]\;, \non
\d_1\psi_1& =-\eta_1 N\D_+[\D_-^2,~\vph]\;, &  \d_1\psi_2& =-\eta_1\D_+D\;.
\end{align}
In the same way, for $\d_2$ we have
\begin{align}
\d_2\vph& =\eta_2\D_-\psi_2\;, &  \d_2D& =-\eta_2 N\D_-[\D_+^2,~\psi_1]\;, \non
\d_2\psi_1& =\eta_2\D_-D\;, & \d_2\psi_2& =-\eta_2 N\D_-[\D_+^2,~\vph]\;.
\end{align}
It can be easily verified that these susy transformations form a closed algebra. 
For instance if
\begin{align}
\d_2\Phi=\hat{\eta}_2\hat{\D}_-[Q_2,~\Phi]\;, &&
\d'_2\Phi=\hat{\eta}'_2\hat{\D}_-[Q_2,~\Phi]\;,
\end{align}
we have \beq
(\d_2\d'_2-\d'_2\d_2)\Phi=2\eta'_2\eta_2\hat{\D}^2_-\left\{Q_2,~[Q_2~\Phi]
\right\}=2\eta'_2\eta_2\hat{\D}^2_-[\hat{\D}^2_+,~\Phi]\;.
\eeq
Likewise for $\d_1$:
\beq
(\d_1\d'_1-\d'_1\d_1)\Phi=2\eta'_1\eta_1\hat{\D}^2_+[\hat{\D}^2_-,~\Phi]\;,
\eeq
and \beq (\d_1\d_2-\d_2\d_1)\Phi=0\;. \eeq

\section{Why there is no inconsistency}
The supersymmetry variation of a product of two superfields follows a
\emph{modified} Leibniz rule. For instance, if  we consider the variation under
 $Q_1$ we have:
\beq \label{modleib}
\d_1(\Phi_1\Phi_2)=\hat{\eta}_1\hat{\D}_+[Q_1,~\Phi_1\Phi_2]=(\d_1\Phi_1)
\Phi_2+(\hat{\D}_+\Phi_1\hat{\D}_-)\d_1\Phi_2\;,
\eeq
where $\hat{\D}_+\Phi_1\hat{\D}_-$ is
a shifted field. Similar expression can be obtained for variations under
$Q_2$ and for translations. Let us denote by $\Phi|_0\equiv\hat{\vph}$ the first
component of the superfield $\Phi$ in the $\theta$ expansion, namely its
 diagonal part in the matrix representation of eq.~(\ref{phi}). As superfields
 are represented by triangular matrices we have:
\beq
(\Phi_1\Phi_2)|_0=\hat{\vph}_1\hat{\vph}_2=\hat{\vph}_2\hat{\vph}_1=(\Phi_2\Phi_1)|_0\;.
\eeq
Nonetheless, \emph{superfields do  not commute}:
\beq
\Phi_1\Phi_2\ne\Phi_2\Phi_1\;.
\eeq
Let us go back to eq.~(\ref{modleib}) and take the diagonal part
(i.e. first component) of both terms. We get
\begin{align}
\d_1(\Phi_1\Phi_2)|_0& =\d_1(\hat{\vph}_1\hat{\vph}_2)=
\hat{\eta}_1\hat{\D}_+(\hat{\psi}^{(1)}_1\hat{\vph}_2+\hat{\vph}_1\hat{\psi}^{(2)}_1)\;,
\label{con1} \\
\d_1(\Phi_2\Phi_1)|_0& =\d_1(\hat{\vph}_2\hat{\vph}_1)=
\hat{\eta}_1\hat{\D}_+(\hat{\psi}^{(2)}_1\hat{\vph}_1+\hat{\vph}_2\hat{\psi}^{(1)}_1)\;. 
\label{con2}
\end{align}
In \cite{Bruckmann:2006ub} it is claimed that equations (\ref{con1}) and (\ref{con2}) constitute a 
contradiction because
their r.h.s.\ are different whereas the l.h.s.\ happen to coincide. In fact there 
is no contradiction. In eq.~(\ref{con1}) $\hat{\vph}_1\hat{\vph}_2$ is regarded 
as first component
(diagonal part) of $\Phi_1\Phi_2$, while in eq.~(\ref{con2})
$\hat{\vph}_2\hat{\vph}_1$  ($= \hat{\vph}_1\hat{\vph}_2$) is
regarded as first component of $\Phi_2\Phi_1$. As a matter of
fact, in a supersymmetric field theory \emph{only transformations
of superfields are well defined. Transformations of single
components are well defined only if all the components of the
superfield are specified.}
In particular if the action is given in terms of superfields its variation with
respect to supersymmetry transformations is unambiguously determined.

\section{The action}
The usual action of supersymmetric quantum mechanics is given by
\beq \label{contact} S_{cont}=\int \ud
x\,\ud\t_1\,\ud\t_2\left[\frac{1}{2}\cd_2\Phi\cd_1\Phi+\mathrm{i}F(\Phi)\right]
\eeq
where $F(\Phi)$ is a superpotential.
Here our aim is to construct a matrix (lattice) action which
reproduces (\ref{contact}) in the continuum limit. First of all we
need a matrix representation for the covariant derivatives.
Following the same reasoning which led to matrix supercharges
(\ref{superc}) we define
\begin{align}
\cd_1=\frac{\prt}{\prt\t_1}-N \t_1\hat{\D}^2_-\;, &&
\cd_2=\frac{\prt}{\prt\t_2}-N \t_2\hat{\D}^2_+\;.
\end{align}
It is straightforward matrix algebra to check that these covariant
derivatives anticommute with all supercharges
\begin{align} \label{dq}
\{\cd_i,~Q_j\}=0\quad\forall~i,~j & & \text{and also} & &
\{\cd_1,~\cd_2\}=0\;.
\end{align}
Besides we have
\begin{align}
\cd_1^2=-N \hat{\D}^2_-\;, && \cd_2^2=-N \hat{\D}^2_+\;.
\end{align}
A suitable candidate for our matrix action is the following \beq
\label{latact}
S=\Tr\left(\left\{\frac{\prt}{\prt\t_2},~\left[\frac{\prt}{\prt\t_1},
~\frac{1}{2}[\cd_1,~\Phi]
[\cd_2,~\Phi]+\mathrm{i}F(\Phi)\right]\right\}\right) \eeq which
for the kinetic reads in terms of the component fields: \beq
\label{kin1} S_{kin}\propto\Tr\big( -N
\psi_1[\D^2_+,~\psi_1]-D^2-N^2 [\D^2_-,~\vph][\D^2_+,~\vph]+
 N [\D^2_-,~\psi_2]\psi_2 \big)\;. \eeq
 The invariance of (\ref{latact}) under supersymmetry
 transformations can be easily proved. First we notice that
 in (\ref{latact}) the derivatives with respect to $\t_i$ can be
 replaced by the corresponding $Q_i$ without affecting the trace.
 We can write then \beq
\label{latact2}
S=\Tr\left(\left\{Q_2,~\left[Q_1,~\frac{1}{2}\Psi_1
\Psi_2+\mathrm{i}F(\Phi)\right]\right\}\right) \eeq where we have
defined the fermionic superfields
\begin{align}
\Psi_1=[\cd_1,~\Phi]\;, && \Psi_2=[\cd_2,~\Phi]
\end{align}
that satisfy the orbifold  condition. Let us consider now the
variation $\d_1 S$ of the action defined according to
eq.~(\ref{d1d2}): \beq \label{d1s}\d_1 S =
\Tr\left(\left\{Q_2,~\left[Q_1,\hat{\eta}_1\hat{\D}_+[Q_1,~\frac{1}{2}\Psi_1
\Psi_2+\mathrm{i}F(\Phi)]\right]\right\}\right)\;. \eeq By using
Jacobi identities and eq.s (\ref{2lat}) it is easily seen that the
expression under trace in (\ref{d1s}) is a commutator of
$\hat{\D}^2$ with something and hence it vanishes when the trace
is taken. The invariance of the action is then proved in complete
generality.

\section{Conclusions}

We have shown, in the simple one dimensional example of $N=2$
supersymmetric quantum mechanics, that supersymmetry
transformations on the lattice can be defined without any
ambiguity with the aid of the modified Leibniz rule if the
superfield formalism is consistently used. We expect that the
situation of higher dimensional models will be similar. We already
showed in \cite{D'Adda:2005zk} that the $N=2$ supersymmetric Yang-Mills theory
in two dimensions can be formulated on the lattice in a way that
the action is exact with respect to the four nilpotent
supersymmetry charges, thus ensuring exact supersymmetry under all
of them. We also showed that $N=4$ supersymmetric Yang-Mills 
theory in three dimensions can be formulated on the lattice 
in a similar way \cite{DKKN3}.

There is however a difference between supersymmetry and, for
instance,  translational invariance on the lattice. Supersymmetry
transformations are defined through eq.s (\ref{d1d2}) \emph{and}
the modified Leibniz rules. However, at least to our
understanding, the latter are not the result, as in the case of
translations, of a well defined transformation on the fields (or
superfields) as in (\ref{transl}). This is related to the specific
nature of supersymmetry and to the vanishing of powers higher than
one of the supersymmetry parameters $\eta_i$, which makes it
difficult to conceive how the modified Leibniz rules could result
from the higher orders in the supersymmetry variation. Some new
idea might be necessary to achieve that, and work is in progress
towards that aim. The situation may be summarized by saying that
we have the exact symmetry, but not the corresponding field
transformations. On one hand this suggests  that the problem is
not fully understood yet, as already mentioned, and on the other
hand is not without consequences as it appears problematic to
write exact Ward identities without knowing the underlying
symmetry transformations of the (super)fields. Work is in progress
in this direction too.

\end{document}